\documentclass{article}

\usepackage[final]{neurips_2022}

\usepackage[utf8]{inputenc} % allow utf-8 input
\usepackage[T1]{fontenc}    % use 8-bit T1 fonts
\usepackage{hyperref}       % hyperlinks
\usepackage{url}            % simple URL typesetting
\usepackage{booktabs}       % professional-quality tables
\usepackage{amsfonts}       % blackboard math symbols
\usepackage{nicefrac}       % compact symbols for 1/2, etc.
\usepackage{microtype}      % microtypography
\usepackage{xcolor}         % colors
\usepackage{graphicx}         % colors

\title{Quantifying attention via dwell time and engagement in a social media browsing environment}

\author{%
  Hause Lin\thanks{These authors contributed equally.} \\
  Levene Graduate School of Business\\
  University of Regina\\
  Regina, CA S4S 0A2 \\
  \texttt{hle952@uregina.ca} \\
   \And
  Ziv Epstein$^*$ \\
  The Media Laboratory\\
  Massachusetts Institute of Technology\\
  Cambridge, MA 02143 \\
  \texttt{zive@mit.edu} \\
   \And
  Gordon Pennycook \\
  Levene Graduate School of Business\\
  University of Regina\\
  Regina, CA S4S 0A2 \\
  \texttt{gordon.pennycook@uregina.ca} \\
   \And
  David Rand \\
  Sloan School of Management\\
  Massachusetts Institute of Technology\\
  Cambridge, MA 02143 \\
  \texttt{drand@mit.edu} \\
}

\begin{document}

\maketitle

\begin{abstract}
Modern computational systems have an unprecedented ability to detect, leverage and influence human attention. Prior work identified user engagement and dwell time as two key metrics of attention in digital environments, but these metrics have yet to be integrated into a unified model that can advance the theory and practice of digital attention. We draw on work from cognitive science, digital advertising, and AI to propose a two-stage model of attention for social media environments that disentangles engagement and dwell. In an online experiment, we show that attention operates differently in these two stages and find clear evidence of dissociation: when dwelling on posts (Stage 1), users attend more to sensational than credible content, but when deciding whether to engage with content (Stage 2), users attend more to credible than sensational content. These findings have implications for the design and development of computational systems that measure and model human attention, such as newsfeed algorithms on social media. 
\end{abstract}

% https://attention-learning-workshop.github.io/
% We invite you to submit papers (up to 9 pages for long papers and up to 5 pages for short papers, excluding references and appendix) in the NeurIPS 2022 format. All submissions will be managed through OpenReview submission website. The final submission including main paper, references and appendix should not exceed 12 pages. Supplementary Materials uploads are to only be used optionally for extra videos/code/data/figures and should be uploaded separately in the submission website.

% Attention in human-computer interaction
%     How do we detect aspects of human attention during interactions, from sensing to processing to representations?
%     What systems benefit from human attention modeling, and how do they use these models?
%     How can systems influence a user’s attention, and what systems benefit from this capability?
%     How can a system communicate or simulate its own attention (humanlike or algorithmic) in an interaction, and to what benefit?
%https://attention-learning-workshop.github.io/

\section{Introduction}
In our current attention economy \citep{wu2017attentionmerchant}, digital ecosystems and social media environments are designed to grab and vie for users' attention. Social media platforms often leverage vulnerabilities in human psychology to distract users and exploit their attention \citep{lorenzspreen2020}, leading many to argue that the ways digital platforms quantify and extract value from users' attention have led to a crisis in attention \citep{hwang2020subprime, wu2017attentionmerchant}. 

However, it remains unclear how attention actually operates in digital ecosystems. Indeed, understanding how attention can be detected, modeled and influenced by computational systems is crucial for promoting better digital ecosystems. In this paper, we draw on research and methods from cognitive science, psychology, AI, and human-computer interaction to propose and validate a model for attention for digital ecosystems. 

\subsection{Measuring and quantifying attention online} 
Decades of attention research has led to many insights into how attention operates in many contexts \citep{buschman2015attention, simon1971}. Digital environments such as social media offer yet another context that requires new ways of measuring, quantifying, and understanding attention (see \cite{lorenzspreen2020}). The digital advertising industry is among the first to systematically measure and quantify attention online \citep{hwang2020subprime}. By standardizing user engagement and attention with metrics like the number of clicks and dwell time, it turned advertising into an online marketplace where user attention is commodified and traded via real-time bidding \citep{hwang2020subprime, wang2016advertise}, and these attention metrics have also been used to predict purchase intentions and behaviors of (even anonymous) website visitors \citep{mokryn2019will}.

Although researchers in different fields such as collaborative filtering and information retrieval have long recognized the value of quantifying attention and engagement via dwell time \citep{resnick1994grouplens}, it is only in the last decade where work has been done to measure and model dwell time for different types of digital content \citep{lamba2019modeling}, and use it determine whether digital content is useful for and relevant to individual users \cite{liu2011using, yi2014beyond}. Surprisingly, little work has has been done to integrate dwell time and engagement data, which provide different yet complementary measures of attention. As far as we are aware, the only work is by \cite{lagun2016understanding}, who proposed a four-level taxonomy (bounce [<10s dwell time], shallow engagement, deep engagement, complete engagement [dwell and interact]). However, because this taxonomy does not offer details into whether and how attention operates differently across levels, it does not provide insights into how to optimize different types of attention.

\begin{figure}[h]
    \centering
    \includegraphics[width=\textwidth]{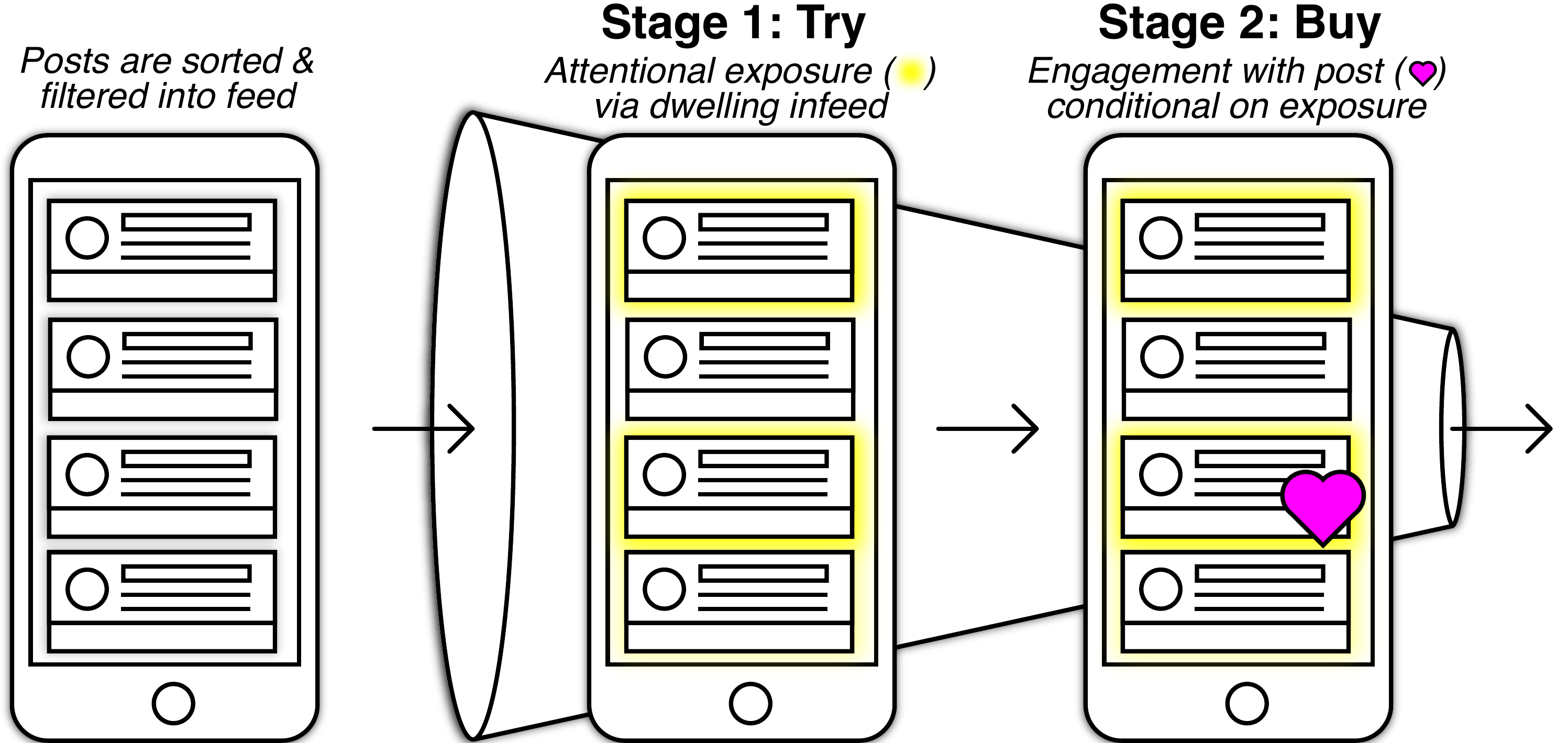}
    \caption{Two-stage model of attention in social media environments (Try + Buy).}
    \label{fig:trybuymodel}
\end{figure}

\subsection{Understanding and dissociating attention with a two-stage try-buy model}
Here, we propose a two-stage model (Try + Buy) that integrates different ways of conceptualizing attention in social media environments (Figure \ref{fig:trybuymodel}). Users are initially exposed to content in an algorithmically-generated newsfeed (Stage 1), and then engage with content conditional on having been exposed to it (Stage 2). Crucially, our model jointly considers distinct attention dynamics at two different stages. First, the extent to which users attend to a piece of content reflects the amount of ``trying'' (Stage 1), which can be quantified continuously via dwell time. Second, engagement behavior such as sharing or liking content reflects ``buying'' (Stage 2). 

Previous work focused largely on Stage 2---what causes people to engage with or ``buy'' content in digital ecosystems and what are the consequences \citep{chen2021share, salganik2006experimental}. However, some work highlights how people have to first sample or ``try'' content before they decide whether to ``buy'' it \citep{krumme2012quantifying, van2016align}, and crucially, different processes like social influence may operate differently in the ``try'' versus ``buy'' stages (see also \citet{epstein2021social} for a review). Crucially, unlike previous work that used two-stage models to primarily define and measure content quality \citep{wu2018beyond, abeliuk2017taming} and jointly predict ``trying'' and ``buying'' \citep{zhou2018jump}, this paper focuses on dissociating the ``try'' and ``buy'' stages and examining how attention operates differently in these stages afforded by systems like social media environments. 

An important implication of our ``try-buy'' model is that systems that optimize for dwell time (e.g., TikTok \citep{tiktok_nyt, tiktok_wsj}) versus engagement (e.g., Facebook \citep{fb_wpo}) focus on different attention dynamics by design, which in turn may lead to different information environments. Thus, in this paper, we use dwell time and engagement data to provide insights into how attention operates in the ``try'' and ``buy'' stages, and reflect on how algorithm designers may optimize for these signals responsibly. 

\section{Methods} \label{method}

\subsection{Dwell time for posts on social media feed}
We recruited a convenience sample of Americans (N=644), of which 628 completed the survey on desktop computers (n=483) or mobile devices (n=145), using the recruitment platform Prolific. (compensation: \$9 USD/hour; total amount: approx. \$1700). Our participants had mean age of 35.7 (46.5\% female, 66\% white). At the start of the survey, participants provided informed consent and were routed to Yourfeed, a website we designed that displays content in a scrolling feed layout \citep{epstein2022yourfeed}. The user interface mirrors the appearance of commonly used social media sites, such as Facebook or Twitter. Participants saw a modal that said ``Thank you for participating! Next, you will see a social media newsfeed, configured just for you. Please browse this newsfeed like you usually would for social media. For each post, indicate whether you would consider sharing it with your network.'' 

Crucially, this platform measured dwell time by considering how much time a participant spent on each post, which was determined based on how much each post was in the visible area of the browser window (also known as viewport time; \citep{lagun2016understanding}). Occasionally, two posts could be fully visible in the browser window---in such cases, we assumed participants were viewing both because it was impossible to determine exactly which post they were looking at. This design detail reflects a trade-off between internal and external validity \citep{Lin2020promises}, and our platform was designed specifically to mimic the user interfaces of existing social media platforms like Twitter and Facebook, which often have two or more posts fully in view.

The website displayed 120 actual and recent social media posts to each participant in a scrollable feed, and participants could click to share or/and like any post (these behaviors were hypothetical engagement decisions, and did not affect what other participants saw). Of the 120 posts shown, 90 were randomly sampled from a set of 200 political and non-political news items \citep{epstein2022many, pennycook2021practical}, half of which are true and half false. The other 30 were randomly sampled from a set of 76 opinion and mundane news items. The mundane posts were sourced from tabloid sites (e.g. The Sun, Daily Mail) and opinion posts are opinion pieces from reputable sources (e.g., New York Times Economist). All posts contained both an image and text (see Appendix \ref{appendix} for example posts).

\subsection{Feature ratings for each post}

In addition to the task described above, we also conducted a separate rating survey to obtain out-of-sample post-level features for each of the 276 posts used. We recruited participants (N=1248) from the recruitment site Lucid to rate these posts (compensation: \$9 USD/hour; total amount: approx. \$1800), and included in our analyses only participants (N=872) who passed two attention checks. After providing informed consent, participants rated 40 randomly selected posts (of 276) on one of eight dimensions: 1) If you were to see the above article on social media, how likely would you be to share it?, 2) Are you familiar with the above headline (have you seen or heard about it before)?, 3) What is the likelihood that the above headline is true?, 4) Assuming the above headline is entirely accurate, how favorable would it be to Democrats versus Republicans?, 5) How provocative/sensational is this headline?, 6) How informative is this headline?, 7) How surprising is this headline?, and 8) How impactful is this headline? 

We computed the mean across participants to compute a single estimate for each post feature (an average of 15.06 ratings per post per feature). We then use these ratings below to examine how they might be associated with the two stages of the try-buy model. Note that true posts were rated as significantly more true than false headlines (b = 1.11, p < 0.001) and were more likely to be shared by participants \citep{epstein2022many}. Moreover, posts participants indicated they were more likely to share in this rating survey were also posts that were shared more frequently by participants in the actual experiment described above (r = 0.26, p < 0.001).

\subsection{Dwell time preprocessing}

Following \cite{lin2022ddm}, we excluded posts whereby dwell times were longer than 30s. We then excluded the first three and last three posts because dwell times could not be determined precisely when participants were reading the instructions at the start or deciding whether to submit and proceed to the next phase of the study at the end of feed. 

Because scrollable social media feeds introduce dependencies between dwell time and engagement (i.e., to engage with a post, people have to slow down and click the share/like buttons; but when they do not want to engage with a post, they do not have to slow down or click any button), we dissociated the motor and attentional components that contribute to dwell time. To do so, we fitted a Bayesian hierarchical mixed-effects model to predict dwell time as a function of the number of times participants engaged with any given post. The participant-specific coefficients provided an estimate of the time it takes for each participant to engage once with a post (``movement time''), and we adjusted dwell times by subtracting ``movement time'' to eliminate the motor component of dwell time, which should leave us with primarily the attentional component. After which, we excluded posts with dwell times shorter than 0.15s because it is unlikely that participants could attend to and evaluate post features so rapidly \citep{lin2022ddm}. Data and code to reproduce the experimental results can be found \href{https://osf.io/duq82/?view_only=4b0e52e6733d43918ce8dc9bafc7d99c}{here}.

\section{Results}

\subsection{Post features and dwell correlations}

To investigate whether dwell times correlated with the eight post features, we computed the mean dwell times (across participants) for each post and correlated them with post features. As shown in Figure \ref{fig:correlations}, the features correlated with each other, and three correlated significantly with dwell (e.g., surprising and true posts had longer and shorter dwells, respectively). Crucially, these correlations suggest dwell captures attention exposure and dynamics and the amount of ``trying.''

\begin{figure}[h]
    \centering
    \includegraphics[width=\textwidth]{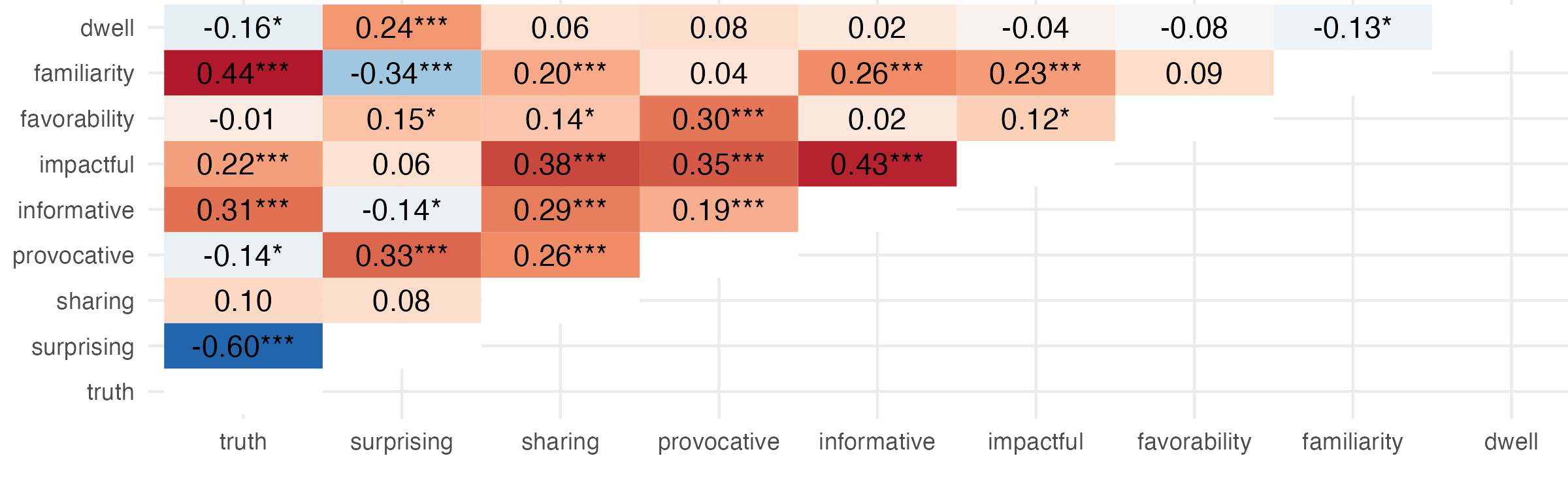}
    \caption{Correlations between post features and dwell.}
    \label{fig:correlations}
\end{figure}

% \begin{table}[ht]
% \caption{Dwell-feature correlations, sorted by magnitude of correlation}
% \centering
% \begin{tabular}{lrrrr}
%   \hline
% Feature & Pearson r & 95\% CI lower & 95\% CI upper & P-value \\ 
%   \hline
% surprising & 0.242 & 0.127 & 0.350 & 0.000 \\ 
%   truth & -0.159 & -0.272 & -0.042 & 0.008 \\ 
%   familiarity & -0.129 & -0.243 & -0.011 & 0.032 \\ 
%   provocative & 0.082 & -0.036 & 0.198 & 0.172 \\ 
%   favorability & -0.081 & -0.197 & 0.037 & 0.179 \\ 
%   sharing & 0.059 & -0.059 & 0.176 & 0.328 \\ 
%   impactful & -0.039 & -0.156 & 0.079 & 0.519 \\ 
%   informative & 0.021 & -0.097 & 0.139 & 0.731 \\ 
%   \hline
% \end{tabular}
% \label{table:dwellcorrelates}
% \end{table}

We then performed principal component analysis (PCA), which revealed that the first two components, together, explained more than half the variance in the data (PC1: 29\%; PC2: 25\%). As shown in Table \ref{table:pca}, relative to PC1 which has large positive weights for the truth, informative, and familiarity features, PC2 has a negative weight for the truth feature and large positive weights for the provocative and surprising features. Thus, PC1 seems to capture variance related to ``credibility,'' whereas PC1 captures variance related to ``sensationalism '' (see Appendix \ref{appendix} for the top posts for each component).

\begin{table}[ht]
\caption{PCA component weights and variance explained} 
\centering
\begin{tabular}{lrrrrrrrr}
  \hline
 & PC1 & PC2 & PC3 & PC4 & PC5 & PC6 & PC7 & PC8 \\ 
  \hline
familiarity & 0.43 & -0.17 & 0.28 & 0.26 & 0.68 & -0.27 & 0.32 & -0.02 \\ 
  favorability & 0.12 & 0.31 & 0.81 & -0.06 & -0.42 & -0.04 & 0.20 & -0.09 \\ 
  impactful & 0.43 & 0.28 & -0.24 & -0.20 & 0.05 & 0.66 & 0.36 & -0.27 \\ 
  informative & 0.45 & 0.07 & -0.35 & -0.40 & -0.29 & -0.63 & 0.15 & -0.00 \\ 
  provocative & 0.17 & 0.52 & 0.12 & -0.29 & 0.38 & -0.03 & -0.67 & 0.01 \\ 
  sharing & 0.35 & 0.28 & -0.19 & 0.79 & -0.28 & -0.04 & -0.23 & -0.00 \\ 
  surprising & -0.26 & 0.54 & -0.13 & 0.05 & 0.14 & -0.06 & 0.41 & 0.66 \\ 
  truth & 0.44 & -0.37 & 0.13 & -0.12 & -0.16 & 0.30 & -0.19 & 0.70 \\ \hline
  variance & 0.29 & 0.25 & 0.12 & 0.09 & 0.08 & 0.07 & 0.06 & 0.04 \\ 
  cumulative variance & 0.29 & 0.54 & 0.66 & 0.75 & 0.83 & 0.90 & 0.96 & 1.00 \\ 
   \hline
\end{tabular}
\label{table:pca}
\end{table}

As with the dwell-feature correlations (Figure \ref{fig:correlations}), PC1 (the ``credibility'' component) correlates negatively (marginally significant) with dwell (r = -0.11, p = 0.063), but PC2 (the ``sensationalism'' component) correlates positively with dwell (r = 0.17 p = 0.005). Together, these correlations suggest that more sensational posts were associated with \emph{more} ``trying,'' but more credible posts were associated with \emph{less} ``trying.'' For a breakdown of the relationships between the component scores and dwell times separately posts that had been engaged with or not, see Figure \ref{fig:correlations}. Given these results, we focus on the two PCA components (instead of the 8 features) in the analyses that follow.

\begin{figure}[h]
    \centering
    \includegraphics[width=\textwidth]{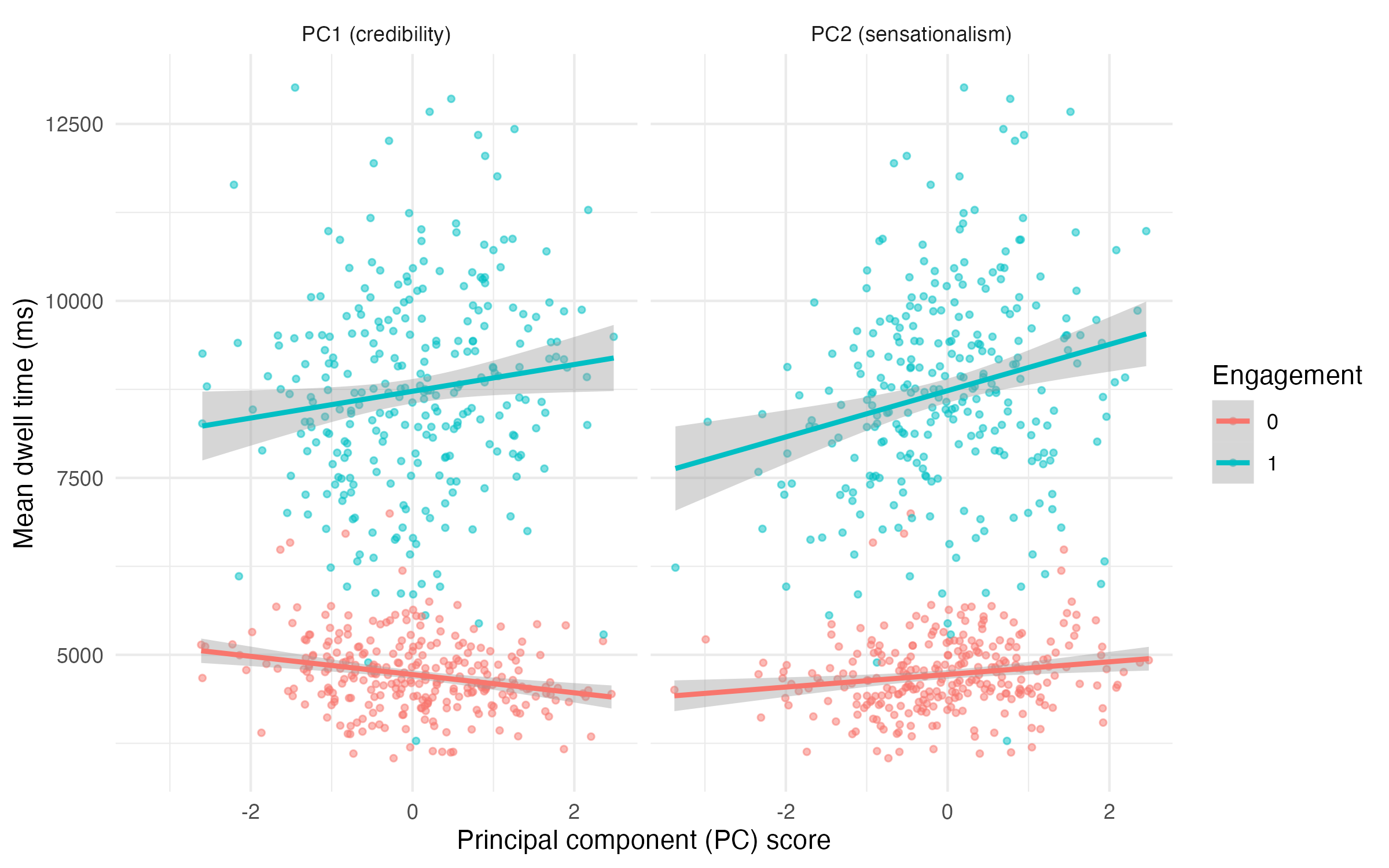}
    \caption{Relationships between principal component scores and dwell for posts that had been engaged with or not. Each dot is one post.}
    \label{fig:correlations}
\end{figure}

\subsection{Evaluating the two-stage model with dwell and engagement analyses}

Next, we examined what influenced the extent to which participants ``tried'' each post by fitting a fixed-effects linear regression to predict dwell time (Table \ref{table:dwellmodel}). For posts participants had engaged with (i.e., shared and/or liked), dwell time was longer (b = 0.31, p < .001). Participants also dwelled longer on posts with higher PC1 (``sensationalism'' component) scores (b = 0.04, p < .001), but less on posts with higher PC2 (``credibility'' component) scores (b = -0.02, p = 0.017). There was also an interaction effect, such participants dwelled even longer on sensational posts they engaged with (b = 0.05, p < .001). Thus, consistent with the results in the previous section, we find that whether participants had engaged with a post and the post's credibility and sensationalism influenced how much participants ``tried'' each post.

\begin{table}[ht]
\caption{Fixed-effect regression predicting log(dwell) as a function of engagement (no: -0.5, yes: 0.5), credibility (PC1 z-scored), and sensationalism (PC2 z-scored)} 
\centering
\begin{tabular}{lrrrr}
  \hline
 & Estimate & SE & t value & Pr($>$$|$t$|$) \\ 
  \hline
engage & 0.311 & 0.025 & 12.348 & 0.000 \\ 
  credibility & -0.017 & 0.007 & -2.405 & 0.017 \\ 
  sensationalism & 0.038 & 0.008 & 4.729 & 0.000 \\ 
  engage:credibility & 0.010 & 0.011 & 0.901 & 0.368 \\ 
  engage:sensationalism & 0.048 & 0.013 & 3.694 & 0.000 \\ 
   \hline
\end{tabular}
\label{table:dwellmodel}
\end{table}
% \vspace{-0.4cm}

Having shown that dwell time captures attentional exposure and dynamics and that it serves as a measure of "trying," we turn to the question of what features are associated with decisions to ``buy'' by fitting a fixed-effect logistic regression to predict engagement (i.e., yes: 1, no: 0). As shown in Table \ref{table:engagemodel}, longer dwell times were associated with an increased probability of engagement (b = 0.36, p < .001). That is, the more participants "tried," the more likely they were to "buy." 

However, in contrast to the model predicting dwell times (i.e., extent of "trying") above (Table \ref{table:dwellmodel}), participants were more likely to engage with credible posts (b = 0.21, p < .001), but less likely to engage with sensational posts (b = -0.22, p < .001). Moreover, dwell interacted with sensationalism, such that sensational posts with longer dwell times were more likely to be engaged with (b = 0.06, p = 0.003). In other words, participants were more and less likely to "buy" credible and sensational posts, respectively. But when they dwelled longer on sensational posts, they were also more likely to then engage with these posts.

\begin{table}[ht]
\caption{Fixed-effect logistic regression predicting engagement as a function of log(dwell) (z-scored), credibility (PC1 z-scored), and sensationalism (PC2 z-scored)} 
\centering
\begin{tabular}{lrrrr}
  \hline
 & Estimate & SE & t value & Pr($>$$|$t$|$) \\ 
  \hline
dwell & 0.355 & 0.029 & 12.355 & 0.000 \\ 
  credibility & 0.212 & 0.049 & 4.361 & 0.000 \\ 
  sensationalism & -0.221 & 0.047 & -4.711 & 0.000 \\ 
  dwell:credibility & 0.011 & 0.020 & 0.538 & 0.590 \\ 
  dwell:sensationalism & 0.062 & 0.021 & 2.921 & 0.003 \\ 
   \hline
\end{tabular}
\label{table:engagemodel}
\end{table}
\vspace{-0.5cm}

% https://attention-learning-workshop.github.io/

\section{Discussion} \label{discussion}
    In this paper, we introduce a two-stage model of attention to conceptualize and understand how attention operates in social media environments. Using an analytic approach informed by our try-buy model, we find dissociations between the ``try'' and ``buy'' stages: in the ``try'' stage, attention (dwell) was focused on sensational posts and not credible posts. Conversely, in the ``buy'' stage, attention (engagement) was focused on credible posts and not sensational posts. However, %it is unclear how our model and findings generalize to environments where users only have the option of ``trying'' or ``buying'' (but not both), and follow-up experiments are required to address this limitation.  In addition, 
    our experiments used data about hypothetical engagement. While past work has shown that self-reported news sharing in surveys correlates with actual sharing on Twitter \citep{mosleh2020self}, future work should replicate these findings with actual engagement data from social media.

Nevertheless, our model and results have important implications for how attention is modeled and leveraged by AI systems in human-computer interactions: For one, algorithmic systems that explicitly optimize for dwell time may prioritize sensational content over credible content and therefore inadvertently proliferate misinformation. Conversely, while optimizing for engagement may indeed surface credible content, we found that people were more likely to engage with sensational content after dwelling more on them, which could create a positive feedback loops that drives the spread of misinformation \citep{hao2021facebook}. Future work is needed to apply our findings to the adaptive dynamics of optimized newsfeed algorithms, and how to align such algorithms with human values by more rigorously evaluating optimization metrics \citep{dmitriev2016measuring}, learning complex multi-variate objectives from stakeholders \citep{stray2021you} and directly giving users control of the algorithms instead of trying to infer their desires \citep{ekstrand2016behaviorism, bhargava2019gobo}.

% \cite{dmitriev2016measuring} propose a method for evaluating metrics for optimization and avoid Goodhart's law. 

% \citet{stray2021you} outlines methodologies for aligning such systems to human values. 
 
% \citet{ekstrand2016behaviorism} argue that explicitly eliciting self-reported preferences, rather than just passively observing what they do, is an important future direction for the design for recommendation systems online. 

% engagement vs. dwell

% Most importantly, how can we use this knowledge to design better online attention systems?

% but our two-stage try-buy model and results highlight differences in these these attention metrics

% limitations

% \cite{lagun2016understanding} propose engagement metrics that capture different levels of engagement: bounce, shallow engagement, deep engagement, and complete engagement. 

% Yourfeed \cite{epstein2022yourfeed, epstein2022many} 

\bibliographystyle{unsrtnat}
\bibliography{references}

\newpage

\appendix

\section{Appendix} \label{appendix}
\subsection*{Interpreting the PCA components}
\begin{figure}[h]
    \centering
    \includegraphics[width=\textwidth]{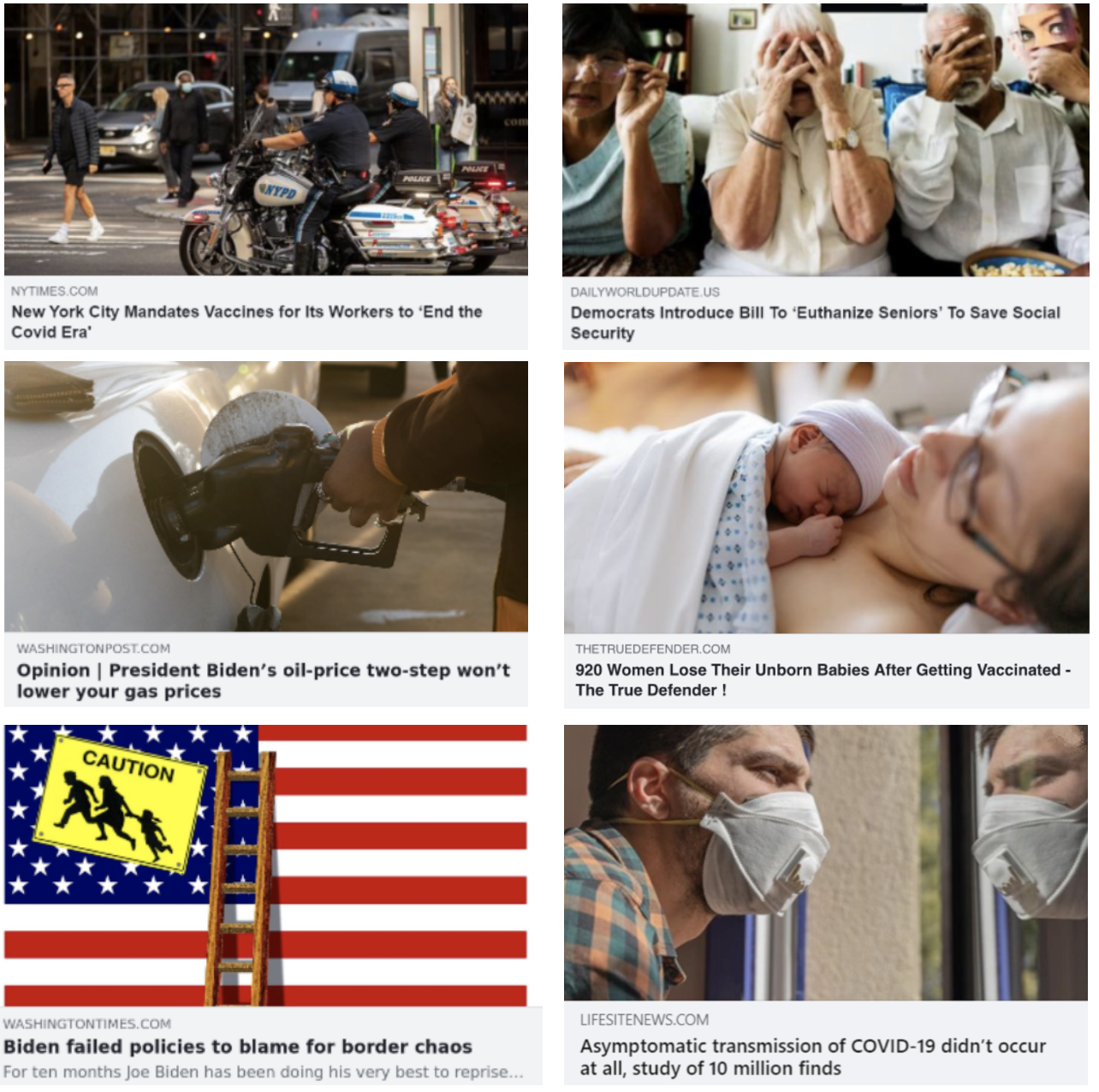}
    \caption{Top headlines for PC1 (left) and PC2 (right).}
    \label{fig:pc1pc2}
\end{figure}

We checked the posts qualitatively (Figure \ref{fig:pc1pc2}), which corroborated our interpretations of PC1 and PC2. For example, the top two PC1 posts were from reputable mainstream new sources: "New York City Mandates Vaccines for Its Workers to `End the COVID Era'" (New York Times) and "President Biden's oil price two-step won't lower your gas prices" (Washington Post). However, the top two PC2 posts were from unreliable or fake news sources: "Democrats Introduce Bill To `Euthanize Seniors' To Save Social Security" (Daily World Update) and "920 Women Lose Their Unborn Babies After Getting Vaccinated" (The True Defender).

\end{document}